\documentclass{ws-spin}
\usepackage{multicol}
\begin{document}

\catchline{1}{1}{2011}{}{}
\markboth{Author's Name}{Paper Title}

\title{MAGNETIC RESPONSE OF Pr$_{1-x}$LaCe$_x$CuO$_4$\\ IN COMPARISON WITH
HOLE-DOPED CUPRATES}

\author{A. Sherman}

\address{Institute of Physics, University of Tartu, Riia 142, 51014 Tartu, Estonia\\
\email{alexei@fi.tartu.ee}}

\maketitle

\begin{history}
\received{Day Month Year}
\revised{Day Month Year}
\end{history}

\begin{abstract}
The origin of differences in the magnetic responses of Pr$_{1-x}$LaCe$_x$CuO$_4$ with $x=0.11-0.12$ and moderately doped $p$-type cuprates is investigated using the $t$-$J$ model, the Mori projection operator technique and dispersions of charge carriers derived from photoemission experiments. These differences are related to the proximity of the former crystal to the boundary of the antiferromagnetic phase and to the remoteness of $p$-type compounds from it. This leads to different nesting vectors of the low-frequency equi-energy contours of carrier dispersions in these crystals. The strong nesting with the antiferromagnetic momentum as the nesting vector produces the commensurate low-frequency response and the coexistence of two spin-excitation branches in Pr$_{1-x}$LaCe$_x$CuO$_4$, while incommensurate nesting vectors in $p$-type crystals lead to the incommensurate low-frequency response and the hourglass dispersion of susceptibility maxima.
\end{abstract}


\begin{multicols}{2}
\section{Introduction}
Magnetic responses of Pr$_{1-x}$LaCe$_x$CuO$_4$ with $x=0.11-0.12$ (PLCCO) and moderately doped $p$-type cuprates are essentially different. In PLCCO the low-frequency susceptibility is commensurate, while in the $p$-type cuprates it is incommensurate, heaving peaks at momenta, which differ from the antiferromagnetic (AF) wave vector ${\bf Q}=(\pi ,\pi)$.\cite{Armitage,Fujita11,Wilson} In PLCCO the dispersion of the susceptibility maxima resembles a cone with the apex point at the frequency $\omega=0$ and at the momentum ${\bf Q}$. In $p$-type cuprates this dispersion has the hourglass shape with the waist at ${\bf Q}$ and at the frequency $25-50$~meV.

In this article an attempt is undertaken to elucidate the origin of this difference. As for $p$-type crystals,\cite{Sherman12} the two-dimensional (2D) $t$-$J$ model of Cu-O planes and the Mori projection operator technique\cite{Mori} are used to calculate the magnetic susceptibility of one of the best-studied $n$-type cuprates PLCCO. The use of the $t$-$J$ model allows one to take proper account of strong electron correlations inherent in both types of cuprates. At the electron concentrations $x=0.11-0.12$ the crystal is near the boundary of the antiferromagnetic phase. In consequence of this the low-frequency equi-energy contours of the electron dispersion are nested with the AF momentum as the nesting vector, which is also termed the band folding.\cite{Armitage} This fact plays a central role in the formation of the commensurate low-frequency response. Besides, the strong nesting leads also to the appearance of a supplementary spin-excitation
branch. As a result the dispersion of the susceptibility maxima has the
shape of a cone. Its upper part is formed by the nested into each other
branches of the usual and supplementary spin excitations. The part near the
apex point is determined by the spin-excitation damping, which is sharply
peaked at {\bf Q}. The same two parts can be singled out in the
dispersion of $p$-type cuprates, with the difference that the supplementary
spin excitations are lacking there and the low-frequency spin-excitation
damping peaks at incommensurate momenta. Together with the damping, the
low-frequency susceptibility splits into four incommensurate maxima.

\section{Main formulas }
In the Mori projection operator technique, the following expression for the
susceptibility of the $t$-$J$ model can be obtained:\cite{Sherman12,Sherman13}
\begin{eqnarray}
\chi({\bf k}\omega)&=&-\frac{h_{\bf k}}{\omega^2-\omega\Pi({\bf k}\omega)-\omega_{\bf k}^2},\label{chi}\\
\omega^2_{\bf k}&=&\bar{\omega}^2\left(1-\gamma_{\bf k}\right)\left(\delta+1+\gamma_{\bf k}\right), \label{wk}\\
\Pi({\bf k}\omega)&=&\frac{Z^2}{8Nh_{\bf k}}\sum_{\bf q}\sum_{\tau\tau'}f({\bf kq})\nonumber\\
&&\times\frac{({\cal E}_{\bf k+q}+\tau\varepsilon^-_{\bf k+q})({\cal E}_{\bf q}+\tau'\varepsilon^-_{\bf q})}{{\cal E}_{\bf k+q}{\cal E}_{\bf q}E_{\bf k+q,\tau}E_{\bf q\tau'}(E_{\bf k+q,\tau}+E_{\bf q\tau'})}\nonumber\\
&&\times\Big[(E_{\bf k+q,\tau}+\varepsilon^+_{\bf k+q}+\tau{\cal E}_{\bf k+ q})\nonumber\\
&&\quad\times(E_{\bf q\tau'}-\varepsilon^+_{\bf q}-\tau'{\cal E}_{\bf q})
-\Delta_{\bf k+ q}\Delta_{\bf q}\Big]\nonumber\\
&&\times\left(\frac{1}{\omega+E_{\bf k+q,\tau}+E_{\bf q\tau'}+i\eta}\right.\nonumber\\
&&\quad+\left.\frac{1}{\omega-E_{\bf k+q,\tau}-E_{\bf q\tau'}+i\eta}\right) \label{pi},
\end{eqnarray}
where the temperature $T=0$, $\omega_{\bf k}$ and $\Pi({\bf
k}\omega)$ are the frequency and self-energy of spin excitations. Several processes contribute to the self-energy.\cite{Sherman13} However, in the
considered range of parameters the main contribution to it is made
by the conversion of a spin excitation into an electron-hole pair, which is described by Eq.~(\ref{pi}). In
Eqs.~(\ref{chi})--(\ref{pi}), $h_{\bf k}$ and $f({\bf kq})$ are slowly varying
functions of the 2D momenta ${\bf k}$ and ${\bf q}$, $\bar
{\omega}\approx 2J$, $Z=\frac{1}{6}$ is the spectral weight of coherent states, which contribute to the self-energy, $\gamma _{\bf k}
=\frac{1}{2}[\cos (k_x )+\cos (k_y )]$, $\delta =2\cdot
10^{-4}-4\cdot 10^{-3}$ describes a gap in the spin-excitation spectrum (\ref{wk}) at ${\bf Q}$ (this gap is connected with the short-range AF order), $N$ is the number of sites, $\eta =0.3-7$~meV is
the artificial broadening,
\begin{eqnarray}
\varepsilon^\pm_{\bf k}&=&\frac{1}{2}\left(\varepsilon_{\bf k}\pm\varepsilon_{\bf k-Q}\right),\nonumber\\
{\cal E}_{\bf k}&=&\sqrt{(\varepsilon^-_{\bf k})^2+\Delta^2_f},\label{bands}\\
E_{\bf k\tau}&=&\sqrt{\left(\varepsilon^+_{\bf k}+\tau{\cal E}_{\bf k}\right)^2+\Delta^2_{\bf k}}.\nonumber
\end{eqnarray}
The dispersions $\pm E_{\bf k\tau}$, $\tau =\pm 1$, correspond to
electron bands of a crystal with the $d$-wave superconducting gap $\Delta_{\bf k} =\frac{\Delta}{ 2}[\cos (k_x )-\cos (k_y )]$,
$\Delta =4.8-9.6$~meV,\cite{Zhao11} and with the band folding across the AF Brillouin
zone border, which is characterized by the potential $\Delta _f =0.14$~eV.
This potential and parameters of the tight-binding dispersion $\varepsilon_{\bf k}$ were obtained\cite{Das} by fitting PLCCO photoemission data.

\section{Results and discussion }
Let us analyze the expression for $\chi''({\bf k}\omega
)={\rm Im}\chi ({\bf k}\omega )$, which follows from Eq.~(\ref{chi}),
\begin{eqnarray}
&&\chi''({\bf k}\omega)=\nonumber\\
&&\quad\frac{-\omega h_{\bf k}{\rm Im}\Pi({\bf k}\omega)}{\left[\omega^2-\omega{\rm Re}\Pi({\bf k}\omega)-\omega_{\bf k}^2\right]^2+\left[\omega{\rm Im}\Pi({\bf k}\omega)\right]^2}.\label{chi2}
\end{eqnarray}
As mentioned above, in the short-range AF order the spin-excitation spectrum
has a gap $\omega _r$ at ${\bf k=Q}$, which is defined by
the relation $D({\bf Q}\omega_r)=\omega_r^2-\omega_r{\rm Re}\Pi({\bf Q}\omega_r)-\omega_{\bf Q}^2=0$. For $\omega >\omega
_r $, where $D({\bf k}\omega )$ vanishes at some combinations of
$\omega $ and ${\bf k}$, the resonant denominator in Eq.~(\ref{chi2}) determines
the frequency and momentum dependencies of $\chi''({\bf k}\omega
)$. For $\omega <\omega _r $, where the denominator is a slowly varying
function of its arguments, the behavior of $\chi''({\bf k}\omega )$
is determined by the numerator of Eq.~(\ref{chi2}) containing the spin-excitation
damping $\vert{\rm Im}\Pi ({\bf k}\omega )\vert $.

\vspace{3ex}
\begin{figurehere}
\centerline{\psfig{file=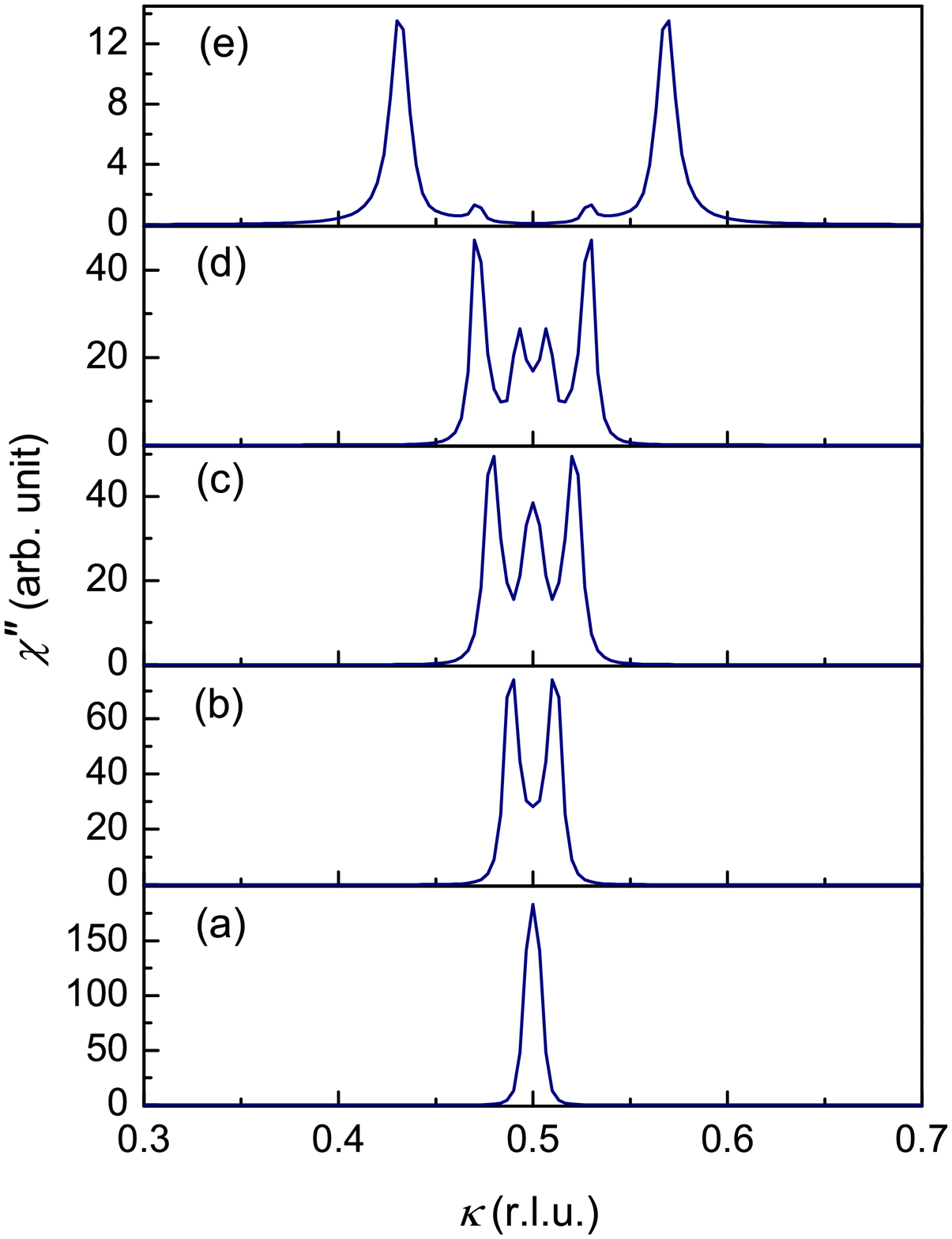,width=6.5cm}}
\caption{The momentum dependence of $\chi''({\bf k}\omega)$ for $\omega=4.8$~meV (a), 16.8~meV (b), 36~meV (c), 48~meV (d) and 108~meV (e). The wave vector varies along a diagonal of the Brillouin zone, ${\bf k}=(\kappa,\kappa)$. $\delta=0.001$ and $\Delta=9.6$~meV.}
\label{Fig1}
\end{figurehere}

\vspace{3ex}
\begin{figurehere}
\centerline{\psfig{file=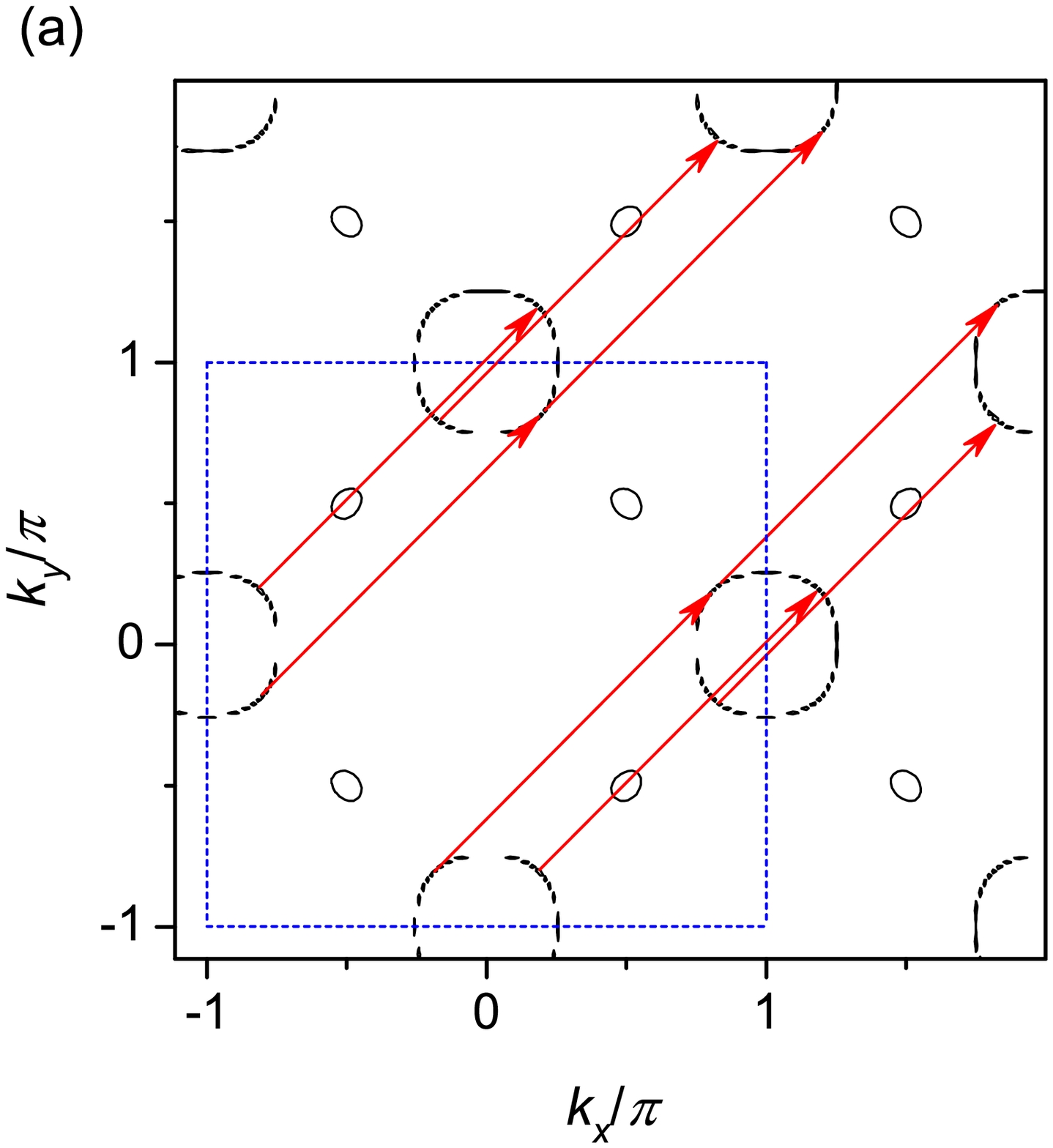,width=6cm}}
\centerline{\psfig{file=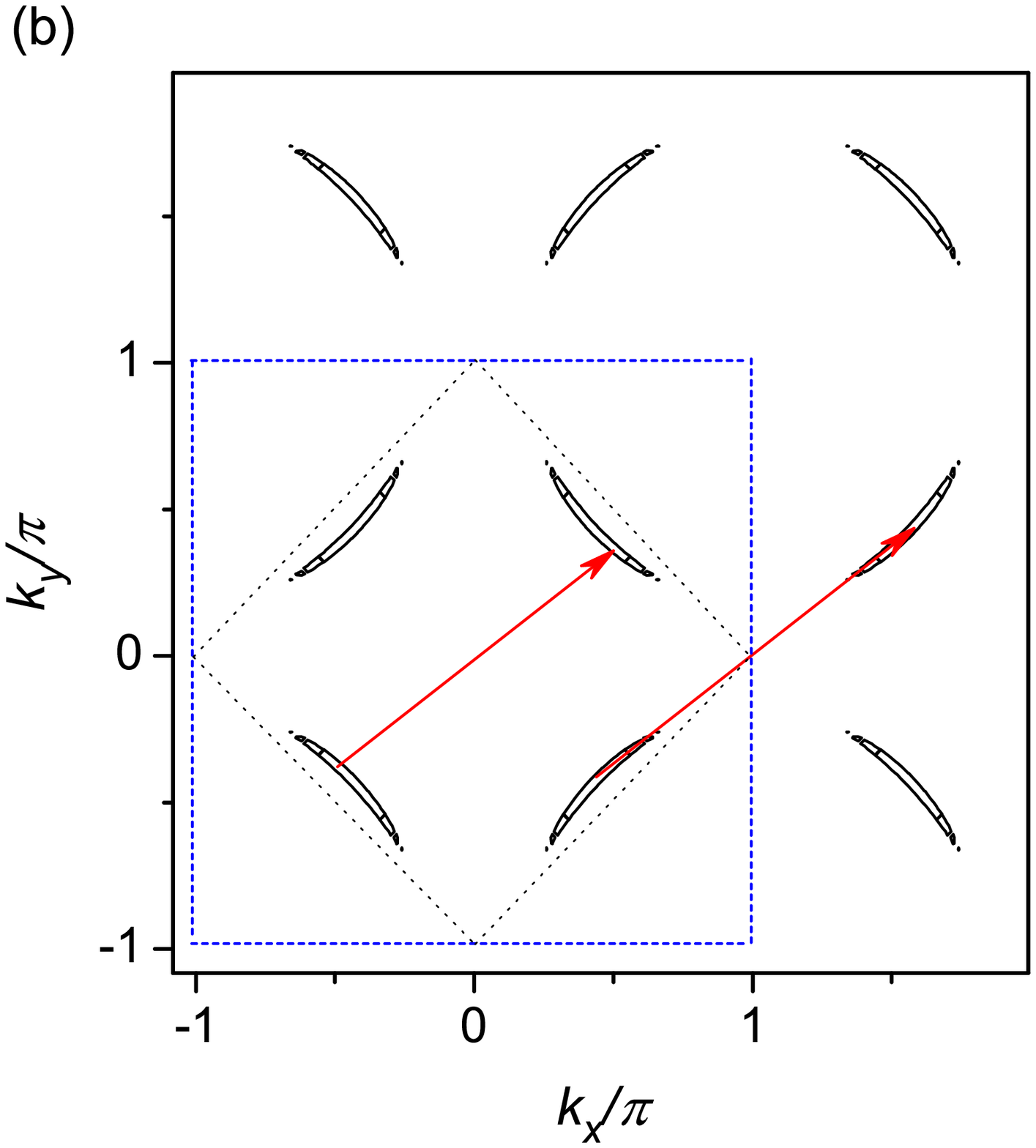,width=6cm}}
\caption{The equi-energy contours of electron dispersion, Eq.~(\protect\ref{bands}), for $\omega
=6$~meV and $\Delta =3.6$~meV (a) and of a typical hole dispersion in
$p$-type cuprates for $\omega =15$~meV and $\Delta =25$~meV (b) (black curves).
The squares shown by blue dashed lines are the first Brillouin zone. Red
arrows show transitions making the main contributions to the spin-excitation
damping.}
\label{Fig2}
\end{figurehere}

The calculated momentum cuts of $\chi''({\bf k}\omega )$ along a
diagonal of the Brillouin zone are shown in Fig.~\ref{Fig1}. For the used parameter
values $\omega _r \approx 6$~meV, and for $\omega \lesssim \omega _r $ the
susceptibility peaks at {\bf Q} [Fig.~\ref{Fig1}(a)] -- the magnetic response is
commensurate. This result is a consequence of the fact that $-{\rm Im}\Pi
({\bf k}\omega )$ in the numerator of Eq.~(\ref{chi2}) peaks sharply at
{\bf Q}. In its turn, this stems from the electron band folding across
the AF Brillouin zone border, which produces nested low-frequency
equi-energy contours with the nesting vector {\bf Q} [see Fig.~\ref{Fig2}(a)].
Thus, in PLCCO the momentum dependence of the low-frequency
susceptibility shows a commensurate maximum with the momentum width growing
with $\omega $. For comparison, in Fig.~\ref{Fig2}(b) low-frequency equi-energy
contours of a $p$-type cuprate are depicted together with the transitions
making the main contribution to the maxima in $-{\rm Im}\Pi ({\bf
k}\omega )$ and $\chi ''({\bf k}\omega )$.
As seen from the
picture, these maxima occur at incommensurate momenta. Indeed, opposite
sides of the magnetic Brillouin zone, shown by black dotted lines in the
figure, are spaced by ${\bf Q}$. The nesting vectors do not coincide
with this momentum. Thus, instead of a single commensurate maximum in
PLCCO, in $p$-type cuprates four incommensurate peaks are observed in the
momentum dependence of $\chi''({\bf k}\omega )$ at low $\omega $.
These peaks produce the down-directed branch of the hourglass dispersion.\cite{Sherman12} Clearly the difference in the nesting vectors in PLCCO and moderately doped $p$-type cuprates is related to the proximity of the former crystal to the boundary of the AF phase and to the remoteness of the latter compounds from this boundary.

\vspace{3ex}
\begin{figurehere}
\centerline{\psfig{file=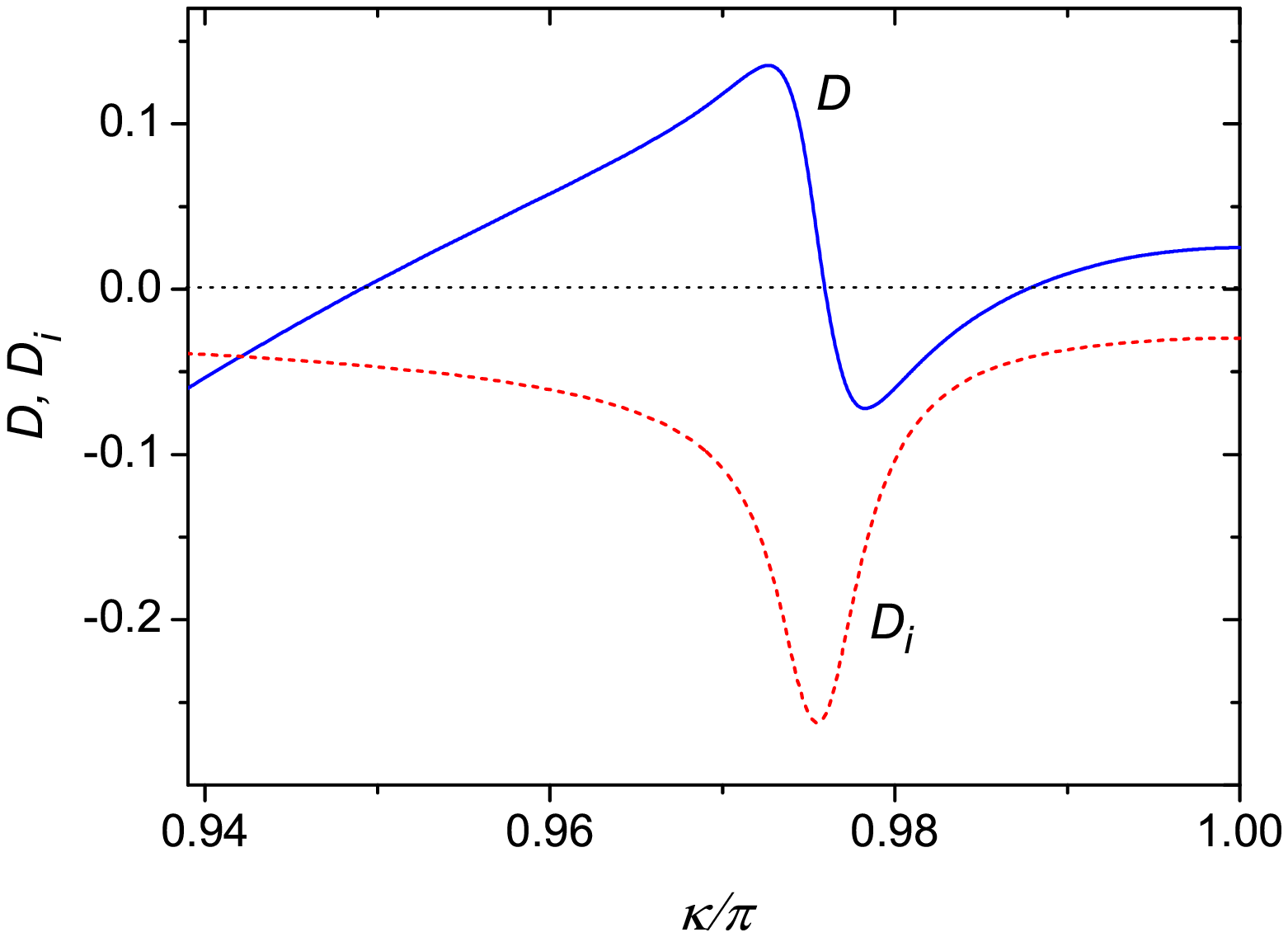,width=7.7cm}}
\caption{The momentum dependencies of the real $D$ (the blue solid line)
and imaginary $D_i=\omega{\rm Im}\Pi({\bf k}\omega)$ (the red dashed line) parts of the denominator in
Eq.~(\protect\ref{chi}) for $\omega=48$~meV, $\delta=0.004$ and $\Delta=9.6$~meV. The wave vector varies along the diagonal of the Brillouin zone, ${\bf k}=(\kappa,\kappa)$.}
\label{Fig3}
\end{figurehere}
For the used parameters as the frequency exceeds 10~meV, the commensurate maximum splits into incommensurate peaks [Figs.~\ref{Fig1}(b)-(e)]. For moderate frequencies $\chi''({\bf k}\omega )$ consists of several closely
spaced peaks. In the experiments,\cite{Fujita11,Wilson} a sole commensurate maximum is observed in the susceptibility up to $\omega\sim 100$~meV. The $k$-width of the maximum grows monotonously with frequency. It is known that samples of PLCCO are inhomogeneous, and apparently the closely spaced peaks of Figs.~\ref{Fig1}(b)-(d) coalesce into the mentioned broad maximum by the inhomogeneity. Only for $\omega\gtrsim 100$~meV, when the internal peaks lose intensity [see Fig.\ref{Fig1}(e)], the maximum is resolved into a nearly circular ridge of peaks around {\bf Q}.\cite{Wilson}

In Fig.~\ref{Fig1}, one spin-excitation branch gives a couple of peaks. The doubled set of peaks in Figs.~\ref{Fig1}(c)-(e) signals that for a fixed
${\omega}$ the equation $D({\bf k}\omega )=0$ has two solutions
corresponding to a small spin-excitation damping. The graphical solution of this equation is shown in Fig.~\ref{Fig3}, which demonstrates three zeros, two of which fall into regions of the small damping $\vert\omega{\rm Im}\Pi({\bf k}\omega)\vert$. Thus, in contrast to $p$-type cuprates, which have one branch of spin excitations, PLCCO has two branches. The branch with the frequency $\omega\approx\omega_{\bf k}$ is analogous to the branch in $p$-type crystals, having the usual spin-wave dispersion. The supplementary branch with higher frequencies exists in a limited frequency range and is also cone-shaped. The reason for the appearance of this latter branch is the region of anomalous dispersion near $\kappa\approx 0.975\pi$ in Fig.~\ref{Fig3}. This region is another consequence of the strong nesting of the carrier dispersion in PLCCO.

\vspace{3ex}
\begin{figurehere}
\centerline{\psfig{file=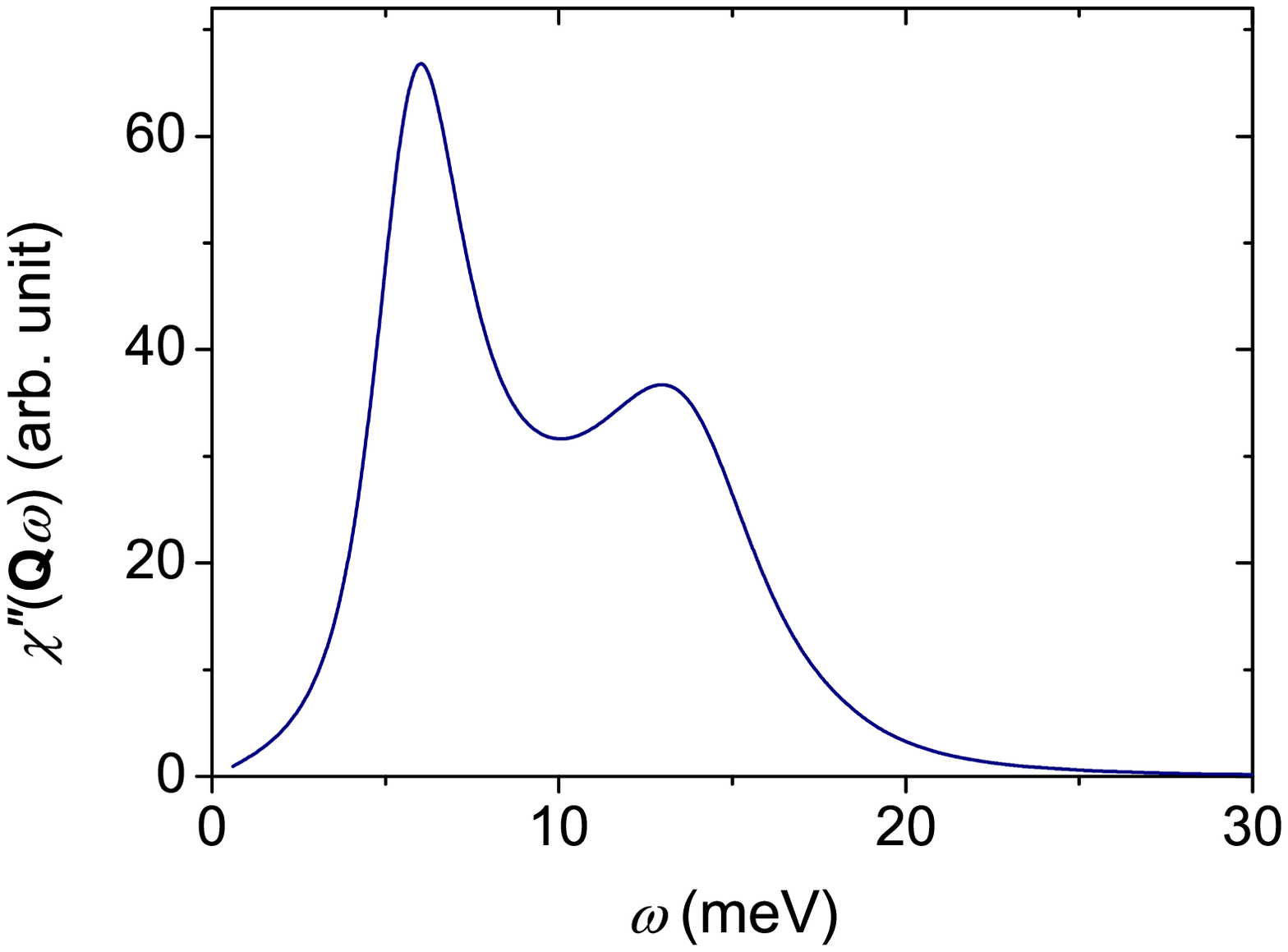,width=7.5cm}}
\caption{The frequency dependence of the susceptibility for ${\bf
k=Q}$, $\delta =5\cdot 10^{-4}$ and $\Delta =4.8$~meV.}
\label{Fig4}
\end{figurehere}

Hence the dispersion of maxima in the susceptibility of this crystal resembles a cone with the apex at $\omega=0$ and ${\bf k=Q}$. The part of this cone for $\omega<\omega_r$ stems from the spin-excitation damping, which is sharply peaked at ${\bf Q}$ for low $\omega$. The part for $\omega>\omega_r$ reflects the spin-excitation dispersions of two nested into each other branches, which have the spin-wave shape with gaps at ${\bf Q}$. In an inhomogeneous crystal, for moderate frequencies the closely spaced maxima of the two branches are apparently merged together into a single broad maximum with a $k$-width growing with $\omega$. For larger frequencies this maximum splits into the circular ridge of maxima with the spin-wave dispersion. This picture is in sharp contrast with the behavior of the susceptibility maxima in moderately doped $p$-type cuprates, where the region $\omega<\omega_r$ is formed by the incommensurate peaks of the spin-excitation damping and looks like a down-directed cone, while the part $\omega>\omega_r$ reflects the dispersion of a single spin-excitation branch with the gap $\omega_r$ at ${\bf Q}$. This results in the hourglass dispersion with the waist at $\omega_r$, which is proportional to the hole concentration.

In the frequency dependence of the susceptibility, two spin-excitation branches of PLCCO have to reveal themselves in two maxima, frequencies of which are close to the gap magnitudes of the branches for ${\bf k=Q}$. An example of such a frequency dependence, which was calculated for the parameters of PLCCO, is shown in Fig.~\ref{Fig4}. It is interesting that a qualitatively similar spectrum with two maxima was recently observed in Ref.~\refcite{Zhao11}. It is worth noting that, together with main maxima, shoulders or weaker maxima at higher frequencies can be disclosed also in earlier measurements of the frequency dependence of the susceptibility (see, e.g., Refs.~\refcite{Armitage,Fujita11,Zhao}). However, a large scatter in the data or a narrow frequency range did not allow the authors of these experiments to draw the conclusion that the spectrum has two maxima.

\section{Concluding remarks }
In summary, the $t$-$J$ model, Mori formalism and the inclusion of the electron
band folding allowed us to interpret the commensurate low-frequency
response, the cone-shaped dispersion of the susceptibility maxima and the
appearance of the second peak in the frequency dependence of the magnetic
susceptibility in PLCCO. Besides, this approach gives an insight
into the origin of the differences in the magnetic responses of this crystal and
$p$-type cuprates.

\section*{Acknowledgments}
This work was supported by the European Regional Development Fund (project
TK114) and by the Estonian Scientific Foundation (grant ETF9371).

\end{multicols}
\end{document}